# RSAED: Robust and Secure Aggregation of Encrypted Data in wireless sensor networks

## Merad Boudia Omar Rafik[1] and Feham Mohamed[1]

[1]STIC Laboratory, Department of Telecommunication, University of Tlemcen, Algeria

om_meradboudia@mail.univ-tlemcen.dz, m_feham@mail.univ-tlemcen.dz

## ABSTRACT

*Recently, secure in-network aggregation in wireless sensor networks becomes a challenge issue, there is an extensive research on this area due to the large number of applications where the sensors are deployed and the security needs. In the last few years, aggregation of encrypted data has been proposed in order to maintain secrecy between the sensors and the sink, so the end-to-end data confidentiality is provided. However, the data integrity was not addressed. In this paper, we propose RSAED that allows integrity verification at intermediate nodes, ensures the base station to receive ciphertexts which come only from legitimate nodes and also improves the efficiency. Through implementation results, we evaluate our scheme using computation and communication overhead.*

## 1. INTRODUCTION

Wireless sensor networks (WSNs) are composed of a large number of sensor nodes, which actually have a wide range of applications such as military surveillance, environmental monitoring and construction safety. Due to their design, sensor nodes tend to have a limited storage space, energy supply and communication bandwidth, and every possible solution that aims to reducing the usage of these resources is widely sought [1]. Data aggregation is one of solutions, which is widely used, it allows in-network processing which leads to lesser packet transmission and reduces redundancy, and hence, helps in increasing the network lifetime [2].

In-network processing is done at aggregator node or intermediate node in the case of multi-hop network, it aggregates the data coming from its child nodes by performing the aggregation function such as min, max, average, sum etc. and sends the result to the upper level node or sink. However, in hostile environments, the aggregated data should be protected from various types of attacks that can be launched by unauthorized or compromised nodes, and hence, security services such as data confidentiality and data integrity are widely desired for providing security [3].

To cope with the security risks, several studies have been proposed to secure data aggregation in WSNs, and they can be classified into two categories: hop-by-hop security schemes and end-to-end security schemes. In the former, the data is encrypted in each node of the network, and before encrypting, each intermediate node needs to decrypt and aggregate the data, this process not only prevents the secrecy of data, but also, results in an important computation overhead and delays. In order to provide data secrecy, several end-to-end security schemes have been proposed in which the data is concealed end-to-end i.e. the data is encrypted only at sensing nodes and decrypted only at the base station. In these schemes, the intermediate nodes perform the aggregation function over encrypted data without decrypting which leads to lesser computation overhead and provides the end-to-end data confidentiality [4]. However, in these solutions, the homomorphic encryption is used and it is known that this kind of encryption suffers from malleability, in other word, given a ciphertext c, an attacker can easily generate a ciphertext c' in order to deceive the base station by accepting the corresponding m' that is related to the original plaintext m, and without necessarily known to the attacker [5]. Therefore, it is of primordial importance to develop secure in-network aggregation schemes that provide both data confidentiality and data integrity.

In this paper, we present RSAED to solve the above problems and also improve the efficiency and the robustness of WSNs. Our contributions are summarized as follow:

- We introduce the additive homomorphic Elliptic Curve El Gamal (ECEG) to provide the end-to-end confidentiality.

- We propose (S-ECEG) a secure ECEG with respect to active attacks.

- We propose RSAED to reduce the overhead at aggregator nodes.

- We provide implementation results of our scheme using TinyECC library.

The rest of the paper is organized as follow: in the next section, we present our system model and design goals. In section III, we give a secure version of ECEG for end-to-end security with respect to active attacks. In section IV, we present our RSAED. In section V, we review some previous related works. Finally, we end the paper by conclusion.

## 2. SYSTEM MODELS AND DESIGN GOALS

### 2.1. Security model

In order to provide the end-to-end confidentiality, the homomorphic encryption is used, which allows calculation on encrypted data without decrypting and prevents the intermediate nodes to access to the plaintext [6]. An encryption algorithm is accepted to be additively homomorphic if:

$$\text{Enc } (\alpha+\beta) = \text{enc } (\alpha) + \text{enc } (\beta) \tag{1}$$

And multiplicatively homomorphic if:

$$\text{Enc } (\alpha \times \beta) = \text{enc } (\alpha) \times \text{enc } (\beta) \tag{2}$$

Both symmetric and asymmetric encryption can satisfy the homomorphic encryption, but recently [7], those based on symmetric cryptography were cryptanalyzed [8], and therefore, for security reasons, it is preferred to use the Public Key Cryptography (PKC). It is known that using PKC incurs a very high computation overhead, but the recent results showed that elliptic curve cryptography is feasible, and can be implemented in devices that have limited resources [9]. The major advantage of using ECC is that it provides the same security level as that offered by existing PKC schemes with smaller key size 'see table 1'.

| Symmetric Key Size (bits) | RSA and DSA Key Size (bits) | ECC Key Size (bits) |
|---|---|---|
| 80 | 1024 | 160 |
| 112 | 2048 | 224 |
| 128 | 3072 | 256 |
| 192 | 8192 | 384 |
| 256 | 15360 | 512 |

**Table 1.** Comparison of key length [10]

In this work, we use Elliptic Curve El Gamal (ECEG) and its additive homomorphic property described in [11], where the authors studied the PKC candidates for end-to-end security in WSNs, and they showed that ECEG is the most promising scheme. In the following, we give a brief introduction to ECC and then describe the additive homomorphism for ECEG.

➢ *Elliptic curve cryptography*

ECC is a public key cryptography approach based on the algebraic structure of elliptic curves over finite fields [10]. There are two types of finite fields where the elliptic curves are defined: prime fields $F_p$, where p is a large prime number, and binary fields $F_{2m}$. In this work, we are interested in the use of elliptic curves over prime fields E ($F_p$). Let p > 3, then a non-supersingular elliptic curve E over $F_p$ is defined as the solution of (x, y) ∈ $F_p$ x $F_p$ to the cubic equation:

$$y^2 = x^3 + ax + b \bmod p \qquad (3)$$

Where a, b ∈ $F_p$ such that $4a^3 + 27b^2$ ≠ 0 (mod p) together with a special point ∞ called *the point at infinity*, The group of points forms an abelian group with addition operation so that the addition of any two points results in another point on the same curve. The addition operation between two points is defined as follows: Given two points $P_1$ and $P_2$, with the coordinates ($x_1$, $y_1$), ($x_2$, $y_2$), respectively. If $P_1$ ≠ $P_2$ then the addition result of $P_1$ + $P_2$ is $P_3 = (x_3, y_3)$, with:

$$x_3 = \lambda^2 - x_1 - x_2 \bmod p \qquad (4)$$
$$y_3 = \lambda(x_1 - x_3) - y_1 \bmod p \qquad (5)$$

And

$$\lambda = ((y_2 - y_1)/(x_2 - x_1)) \bmod p, \text{ if } P_1 \neq P_2 \text{ (adding)} \qquad (6)$$
$$\lambda = ((3x_1^2 + a)/2y_1) \bmod p, \text{ if } P_1 = P_2 \text{ (doubling)} \qquad (7)$$

Note that the inverse of a point $P_1$ is $-P_1 = (x_1, -y_1)$ and $P_1 + \infty = \infty + P_1 = P_1$, The product Q = k.P of a point P on curve with a scalar k is called *scalar point multiplication* and it is performed by a sequence of point addition and point doubling. The security of all cryptographic protocols based on elliptic curves depends on the Elliptic Curve Discrete Logarithm Problem (ECDLP). The ECDLP can be defined as the problem of finding the scalar k such that Q=kP given Q and P. The reader is referred to [10] for more detail.

➢ *The additive homomorphism for EC El Gamal*

**KeyGen:** Given the domain parameters (a,b,p,G,n,E) of an elliptic curve E over finite field $F_p$ where p is a large prime that satisfy equation (3). Where G is the base point of order n, note that n*G = ∞, the private key x is randomly selected from [1, n-1], the public key is Y=xG, another point on the curve.

**Encryption:** Given the plaintext m and Y, output C

*1. k ∈ [1, n − 1]*
*2. M = map (m)= mG*
*3. C= (R, S) = (kG, kY+mG)*

**Homomorphic operation:** Given $C_1$, $C_2$... $C_n$, output C'

*C' = ($k_1$G, $k_1$Y+$m_1$G)+($k_2$G, $k_2$Y+$m_2$G)+...+($k_n$G, $k_n$Y+$m_n$G)*
*C' = (($k_1$+$k_2$+..$k_n$)G, ($m_1$+$m_2$+$m_n$)G+($k_1$+$k_2$+..$k_n$)Y)*

*Decryption:* Given C' and the private key x, output m

*1. $M = S - xR$*
*2. $m = rmap(M)$*

The map function satisfies the desired additive homomorphic property. However, the reverse mapping function is the shortcoming of this scheme, the reverse function maps a given point M into a plaintext m, and thus, the ECDLP (defined above) on M must be resolved. In the following, we give an illustrative example.

Let consider p=11, a=1, b=6. The elliptic curve E over $F_{11}$ is represented by the equation $y^2 = x^3 + x + 6$ mod 11, all points of this curve are (2,4), (2,7), (3,5), (3,6), (5,2), (5,9), (7,2), (7,9), (8,3), (8,8), (10,2), (10,9), along with a point at infinity $\quad$. The point G= (2, 7) is the generator. The group of points generated by G over E ($F_{11}$) can be calculated using the above equations of addition and doubling formulas. The points are: G=(2,7) 2G=(5,2), 3G=(8,3), 4G=(10,2), 5G=(3,6), 6G=(7,9), 7G=(7,2), 8G=(3,5), 9G=(10,9) 10G=(8,8) 11G=(5,9) 12G=(2,4). So the domain parameters are (1,6,11,G,n,E) where the order n is equal to 13 because nG= $\quad$. Assume the private key x=6, we obtain the public key point Y=xG = 6G = (7,9). We take $m_1$=5 and $m_2$=3. The random number $k_1$ for $m_1$ is 7, $k_2$ for $m_2$ is 4.

M= $M_1 + M_2$ = (3,6) + (8,3) = (3,5)

$C_1$ = ($k_1$G, $M_1$G+$k_1$Y) = (7(2,7),(3,6)+7(7,9) = ((7,2),(3,5))

$C_2$ = ($k_2$G, $M_2$G+$k_2$Y) = (4(2,7),(8,3)+4(7,9) = ((10,2),(2,7))

$C_1+C_2$ = ((7,2),(3,5)) + ((10,2),(2,7))

$\qquad$ = ((7,2)+(10,2), (3,5)+ (2,7))

$\qquad$ = ((5,9),(10,9))

M = S-xR = (10,9) − 6(5,9)

$\qquad$ = (10,9) − (2,7)

$\qquad$ = (10,9) + (2,-7)

$\qquad$ = (10,9) + (2,4)

$\qquad$ = (3,5)

The numbers utilized in the above example are very small in order to simplify the comprehension. However, very large numbers are used in practice [12].

In order to provide integrity, we use Message Authentication Code (MAC). It is clear that the end-to-end integrity is impossible since MAC cannot satisfy the additive homomorphic property:

$$\text{MAC } (\alpha+\beta) \quad \text{MAC } (\alpha) + \text{MAC } (\beta) \qquad (8)$$

In RSAED, hop-by-hop verification is adopted to provide both integrity and source authentication. We use HMAC, the hash-based MAC, and the latter has the same properties as the one way hash function including a key [13].

## 2.2. Network model

We assume that the network is divided into stationary clusters, and each cluster contains n sensors nodes,"see Figure1", each of them shares a unique authentication key with its corresponding Cluster Head (CH). The public key of the base station is preloaded in each node. The CH away from the base station forwards the packet to the nearest CH, and both must have a common authentication key. The base station is assumed to have an unlimited energy and powerful processing capability. The pair wise key establishment is out of scope of this paper.

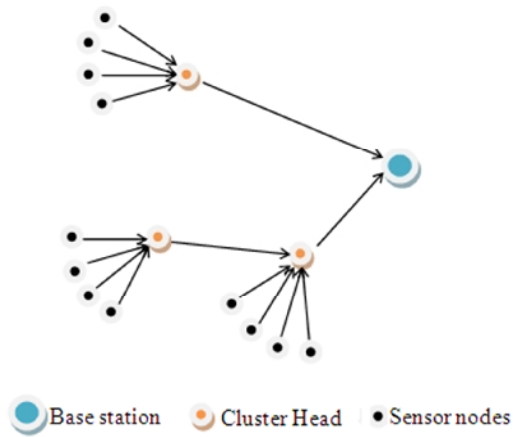

Figure1. Network model

## 2.3. Design goals

**Security**: For dealing the security risks, secure data aggregation scheme must provide the following security requirements.

*Confidentiality:* ensures that the plaintext can only be accessible by the authorized user. All data captured must be encrypted and prevent intermediate node to access to the plaintext.

*Integrity:* ensures that the received data has not been altered, either maliciously or accidentally, during transmission.

*Authenticity:* ensures that the received data is sent by the claimed sender.

*Availability:* ensures the survivability of the network despite denial of service attacks.

*Freshness:* ensures that each message is recent and no old messages replayed by an attacker.

**Efficiency**: a security protocol must be efficient in term of computation and communication overhead in order to preserve energy and prolong the network life time.

**Robustness**: a security mechanism must ensure the availability of packets even with the presence of compromised or malfunctioning nodes.

## 3. SECURE ECEG

In this section, we describe a secure ECEG with respect to active attacks. In [5], the authors studied the security of several homomorphic encryption algorithms, and they showed that ECEG cannot be secure without additional security, because the highest security level it can reach is against passive adversary. In the following, we describe (S-ECEG), a secure ECEG for WSNs.

## 3.1. Description

Once the data is captured by a sensor node, it is encrypted by using the preloaded public key of the base station, and then a tag is produced using the authentication key $k_i^j$ (shared with its corresponding CH [j]), the nonce N and the ciphertext. The nonce is a sequence number that should be used just once for data freshness. Finally, the packet is sent to the CH which includes the encrypted data and the tag. In the following, we describe S-ECEG algorithm for sensor node $SN_i$ which is a member of $CH^j$.

### Algorithm 1: S-ECEG algorithm for sensor $SN_i$

1. Use ECEG to produce $C_i=(R_i, S_i)$
2. Compress the points $R_i, S_i$
3. Compute $tag_i = HMAC(k_i^j, R_i || S_i || N_i)$
4. Send $(R_i, S_i, tag_i)$ to $CH^j$
5. $N_i = N_i + 1$

Once the data received from $SN_i$, $CH^j$ first verifies the $tag_i$ received and then decompress the pair of points. In the other case i.e. if the verification fails then it just drops the packet. This means that the packet was generated, either maliciously or accidentally. Finally, the homomorphic operation is performed using point addition on elliptic curve. In the following, we describe the S-ECEG algorithm for $CH^j$.

### Algorithm 2: S-ECEG algorithm for $CH^j$

1. Compute $tag_i' = HMAC(k_i^j, R_i || S_i || N_i)$
2. Compare $tag_i'$ and $tag_i$; if $tag_i' = tag_i$ then go to 3; otherwise drop the packet.
3. Decompress the points $R_i, S_i$
4. Use the homomorphic property of ECEG to combine valid $C_i$ and produce $C' = (R', S')$
5. Compress the points $R', S'$
6. Compute $tag^j = HMAC(k_j^{bs}, R' || S' || N')$
7. Send $(R', S', tag^j)$ to the sink or the nearest CH.
8. $N'=N'+1$, and also $N_{i,n}=N_{i,n}+1$

This process is repeated until the final encrypted aggregate reaches the base station. Once the data received, the base station first verifies the integrity of all the incoming packets from CHs, and then decompresses the points. Finally, the decryption process is applied using the private key. For retrieving the aggregated plaintext m= $m_1+m_2+ \ldots + m_n$, the ECDLP on M must be resolved and since the base station is assumed to be secure with unlimited available energy, the ECDLP can be calculated efficiently using Pollard-  method on elliptic curve cryptosystem [14].

## 3.2. Security analysis

It is clear that, with the use of a homomorphic encryption algorithm, the data is concealed end-to-end, and therefore the data confidentiality service is ensured. The passive adversaries cannot learn anything from any ciphertext about its corresponding plaintext, thanks to the properties of ECEG algorithm which is probabilistic and where the security relies on the hardness of ECDLP. However, when considering security against active adversaries, a verification of the data integrity is needed in order to ensure that all the data were ported successfully, in S-ECEG, each sensor of the network computes a tag using HMAC algorithm on ciphertext, and every intermediate node then verify the data integrity, execute the homomorphic operation if the verification hold; otherwise, the packet will be dropped, with this process the data integrity of all packet is maintained and all senders are authenticated. In the following, we provide a security analysis of S-ECEG regarding the active attacks described in [5].

***Replay attacks:*** a compromise node cannot replay an old packet due the use of nonce and once the tag is verified, the verifier can be sure that the packet received is recent and not replayed and hence, S-ECEG provides data freshness.

***Malleability:*** the malleability is one of the open research problems in cryptography because the adversary can easily change the ciphertext content, and the fact that this ciphertext is accepted by the reader device once received, so an additional security is severely desired. In S-ECEG, we use HMAC to prevent attacks against data integrity. Therefore, the attacker cannot succeed because it cannot produce a valid tag for a modified ciphertext without the valid key.

***Unauthorized aggregation:*** in S-ECEG, an unauthorized aggregation can be performed if and only if the attacker can obtain the valid MAC key. Otherwise, our scheme is secure against this attack.

***Forge packet:*** due to the properties of public key cryptography, anyone can produce his own ciphertext, so the data integrity is needed, as previously mentioned S-ECEG use HMAC to prevent packet forgery.

***Physical attacks:*** in S-ECEG, if an attacker can physically compromise an aggregator node and can obtain the MAC key, so the data integrity is affected. However, the data confidentiality is maintained due to the use of a public key encryption scheme.

In addition, the forged ciphertext is dropped by intermediate nodes if the tag verification fail, the modified packet can be filtered and dropped as previously as possible, and once the final encrypted aggregate is successfully verified, the base station can be sure that the integrity of all ciphertexts has not been breached and also, the base station can be certain that only legitimate nodes could have produced that ciphertexts. In other words, S-ECEG blocks denial of services attacks, it provides availability of packet even with the presence of compromised nodes in the network. Therefore, our scheme improves the robustness of WSNs.

### 3.3. Performance evaluation and discussion

In this section, we present the performance evaluation of S-ECEG in terms of computation and communication overhead.

- **Computation overhead**

S-ECEG is implemented on MicaZ mote from Crossbow [15], equipped with the 8-bit ATmega128L processor clocked at 7.3728 MHz, 4 kB RAM, and 128 kB Flash memory. NesC as a network embedded system language, using TinyECC library and TinyOS. TinyECC is a freely available ECC implementation library that provides elliptic curve operation over $F_p$, for more details see [16], and TinyOS is an open source operating system for WSN [17], for our application, the SEC recommended secp160r1 parameters are used and the simulation is done using TOSSIM/AVRORA, Tossim as a WSN simulator and Avrora to measure the execution time of our cryptographic functions. Avrora is a cycle–accurate emulator for AVR microcontroller [18], we use the beta 1.7.x version released in November 2007 that support MicaZ platform.

The module SECEG*M.nc* implements initialization, encryption, homomorphic operation and tag calculation functions, namely SECEG.init(), SECEG.encrypt(), SECEG.hom_add() and SECEG.tag() respectively. In Elliptic Curve Cryptography, the most expensive operation is the scalar point multiplication. TinyECC provides a number of optimizations switches, which can be turn on or off based on application's needs, including Sliding Windows Methods (SWM) which improves significantly the execution time of scalar point multiplication especially when the base point is fixed and known a priori. In our code, we use SWM, and HMAC provided by TinyECC. According to [19], the energy consumption P of each arithmetic operation can be calculated by using the formula P = U×I×t, whereby U denotes the voltage, I denotes the current, while the execution time is represented by t, the voltage and the current was assumed to be 3V and 8 mA, respectively. Table 2 summarizes the execution time and energy consumption of various cryptographic function of S-ECEG.

| Operation | Execution time | Energy consumption |
|---|---|---|
| Encryption | 2.844s | 68.256 mj |
| Hom.operation | 1.499s | 35.976 mj |
| Tag calculation | 0.028s | 0.672 mj |

Table 2: Execution time of S-ECEG functions in MicaZ motes.

The encryption involves following ECEG, two scalar point multiplications with a random 160 bits value and another one with the message m, which is taken 8 bits for test, also the points G and Y i.e. the base point and the public key of the base station are known a priori then, by using SWM, a table of precomputed point can be calculated offline and just once. Therefore, a sensor node in our scheme spends 68.937 mj, 68.265 mj for encryption and 0.672mj for tag calculation.

The execution time showed in Table 2 for homomorphic encryption involves the addition of two ciphertexts. In S-ECEG, once the data received, the CH first verifies tags, and for two packets the CH spends 37.992 mj, 2*0.672mj for two tags verification, 35.976mj for homomorphic operation and 0.672 mj for its own tag. In homomorphic operation, the most expensive operation is point decompression. For bandwidth purpose, in S-ECEG we use point compression on curve, and consequently, the CH cannot add ciphertexts in their compressed form, in order to decompress, the CH must resolve the curve equation [20].

- **Communication overhead**

For communication aspect, the ciphertext is represented with a pair of points on curve namely R and S. As previously mentioned, we use point compression which allows representing a point using the minimum possible number of bits where a point is represented by it x coordinate and one additional bit of information. The corresponding y coordinate can be computed using x and this additional bit. In the example presented in section 2.1, assume that $SN_1$ senses the data $m_1$ and $SN_2$ the data $m_2$. The compressed ciphertexts sent to the CH are ((7,0),(3,1)) and ((10,0),(2,1)) for $SN_1$ and $SN_2$, respectively. The additional bit is necessary because there exist two solutions to equation (3) for a given x, for example x=7 have the two solutions y=2 and y=9. In our implementation, we use secp160r1, so the total size of a point is 161 bits (21 bytes). Thus, the size of a ciphertext is 42 bytes. In contrast, the HMAC used output 160 bits (20 bytes) and hence the total size of S-ECEG packet is 62 bytes.

- **Discussion**

The computation overhead at aggregator nodes is very important, and this is due to the expensive operation of elliptic curve, namely the point decompression. By increasing the number of cluster members, aggregator node (CH) needs to perform more and more point decompression over ciphertexts and this leads to energy depletion and consequently, the network lifetime is threatened. Also, the maximum packet size in TinyOS is 39 bytes and for packet transmission in S-ECEG, each node must to send 62 bytes, the solution is to divide the packet into 2 blocks and send each block separately to the aggregator, this solution incurs not only an important overhead but also delay, because the aggregator must to wait the reception of all blocks to perform the homomorphic operation. We can also modify the packet size in TinyOS and send a big packet, but in unreliable links such as in WSNs, this solution increases the bit error rate and decreases the reliability of the network. To overcome the above problems, we propose RSAED in order to reduce the overhead at aggregator nodes. In the proposed scheme, each cluster contains two aggregators instead of one aggregator used in the previous solution.

## 3. RSAED: Robust and Secure Aggregation of Encrypted Data

### 3.1. Scheme details

Our scheme consists of four steps for efficiently and securely aggregates the encrypted data in WSNs. In the following, we describe the scheme.

*1-selection of aggregators*

In this step, two nodes (aggJ1, aggJ2) in each cluster J are dynamically elected 'see figure 2', the election algorithm is based on the available energy of the node and higher energy availability higher the probability to become an aggregator nodes, the two brother aggregators elected must share an authentication key.

*2-data transmission*

Each sensor nodes $SN_i$ computes the ciphertext on the data captured using ECEG algorithm, and sends the two parts (compressed points) of the ciphertext namely R and S to agg1 and agg2, respectively, along with a tag produced by using the key shared with the corresponding aggregator. For example for cluster J:

$$SN_i \rightarrow aggJ1 \qquad ID_{SNi} \mid R_i \mid HMAC\,(K_{i,aggJ1}\,,\,R_i \mid N_i)$$
$$SN_i \rightarrow aggJ2 \qquad ID_{SNi} \mid S_i \mid HMAC\,(K_{i,aggJ2}\,,\,S_i \mid N_i)$$

*3-inter aggregators verification and aggregation*

Once received, the encrypted data is verified by both aggregators and similarly to S-ECEG the packet that failed to the check process will be dropped and the corresponding node id will be maintained in a list. Thereafter, aggJ2 sends a verification message to aggJ1 containing the number of legitimate packets (NLP) and the list of malicious nodes, aggJ1 then compares with its check results and then sends the result to aggJ2, the result is the same packet as received from aggJ2 if the comparison hold; otherwise aggJ1 sends its own check results. For example:

$$AggJ2 \rightarrow AggJ1 \quad ID_{AggJ2} \mid NLP \mid list \mid HMAC\,(K_{aggJ2,aggJ1}\,, ID_{AggJ2} \mid NLP \mid list \mid N_{AggJ2\text{-}J1})$$
$$AggJ1 \rightarrow AggJ2 \quad ID_{AggJ1} \mid NLP \mid list \mid HMAC\,(K_{aggJ1,aggJ2}\,, ID_{AggJ1} \mid NLP \mid list \mid N_{AggJ1\text{-}J2})$$

After the verification process, the brother aggregators filter out all malicious packets and then the addition operation over the parts of ciphertext can be done by first decompress the points and then using points addition on elliptic curve. For example:

$$AggJ1: computes \; R_{aggJ1} = R_1 + \ldots + R_n$$
$$AggJ2: computes \; S_{aggJ2} = S_1 + \ldots + S_n$$

The aggregated points computed are then sent to the base station or the nearest aggregator of the nearest cluster, 'see figure 2'. For example:

$$AggJ1 \rightarrow BS \quad ID_{AggJ1} \mid R_{aggJ1} \mid HMAC\,(K_{aggJ1,bs}\,, R \mid N_{AggJ1\text{-}BS})$$
$$AggJ2 \rightarrow BS \quad ID_{AggJ2} \mid S_{aggJ2} \mid HMAC\,(K_{aggJ2,bs}\,, S \mid N_{AggJ2\text{-}BS})$$

*4-base station verification*

Once the final encrypted aggregate received from aggregators, the base station first verifies the packets integrity and authenticates the senders, and then invokes the decryption process corresponding to ECEG algorithm.

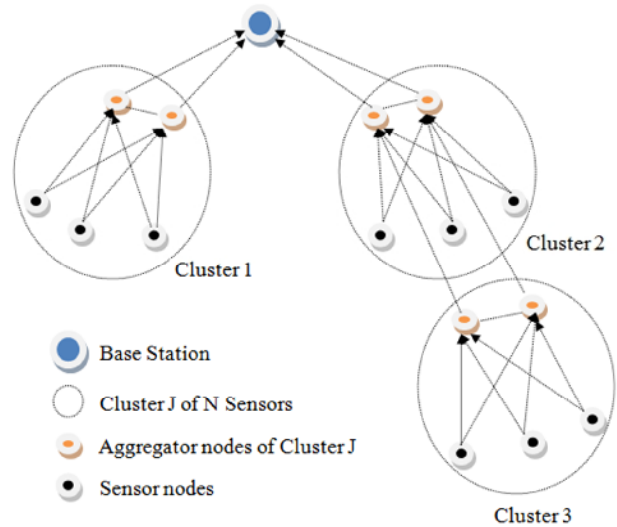

Figure 2. An example of RSAED aggregation model

## 3.2. Security analysis

Our scheme involves a second aggregator node in each cluster, and this aims to reduce the overhead at aggregator nodes of the previous solution, this technique do not affect the security of the scheme. In addition of security analysis of S-ECEG, in the following, we analyze the security of RSAED.

- **Malicious sensor nodes**

In our scheme, there are the two aggregators that perform the role of aggregation over encrypted data, which compute an aggregation result part each, and report it to the base station or the nearest aggregator. The simple way of an attacker to modify the aggregation result and consequently deceive the base station by accepting a forged packet, is to put some malicious nodes into the network and try to inject erroneous packets or even replay previous packets through these nodes. This attack can succeed if the ECEG algorithm is used alone because the scheme is malleable, and only one compromised node is able to modify the correct result. Our scheme allows hop-by-hop verification and since these malicious nodes do not possess the MAC key, they can't disturb the network, and the forged packets are detected and dropped at aggregator nodes. Also, during verification in the third step, the brother aggregators agree on a specific list which contain the same malicious nodes and aggregate the parts (points) which come from the same legitimate nodes.

- **Compromised aggregator nodes**

Usually, the data aggregated by these nodes concerns several nodes of the network, so it is more interesting for an attacker to compromise this kind of sensor instead of a simple leaf sensor measure. Our scheme uses homomorphic encryption and consequently, anything is revealed to the aggregators about the plaintexts. Furthermore, we use hop-by-hop verification that prevents packet forgery and improves the robustness of WSN, for example, if one of the aggregators of a cluster is compromised, and then, detected by the upper level, the corresponding packet is dropped and during the inter aggregators verification of the upper level, the packet received from the brother of compromised node will be also dropped, and hence, transmitting packets which come only from legitimate nodes.

- **Node failure**

In RSAED, the hop-by-hop verification is adopted and can tolerate the node failure. The legitimate packets are available and can reach the base station even with the presence of malfunctioning nodes in the network, and this is done by just skipping the point of failure, and consequently, the robustness of WSN is improved.

## 3.2. Performance evaluation

Our scheme aims to reduce the overhead at aggregator nodes by involving two aggregators in each cluster and splitting the homomorphic operation between them. In the following, we evaluate our scheme in terms of computation and communication overhead.

- **Computation overhead**

An aggregator node in RSAED computes a tag for each received packet and then sends a verification packet to its brother, executes the point decompression and finally the point addition over elliptic curve is performed. The homomorphic operation is calculated in a distributed manner and this decreases the end-to-end delay compared to S-ECEG. The homomorphic operation function of the previous TinyOS application is modified into a lightweight function that takes in input the compressed points and gives in output the addition result of these points. The time taken by this function is about 0.796s and consumes 19.104mj for two points (two parts). Figure 4 shows the effect of increasing the number of sensor nodes on the energy consumption (computation overhead) at aggregator nodes in both RSAED and S-ECEG. It is clear that by using two aggregators reduces significantly the computation overhead at aggregator nodes and hence the network lifetime is improved. Figure 5 illustrates the benefits of using a distributed computing on the end-to-end delay (we assume that the aggregator nodes communicate directly with the base station). For example for 20 sensor nodes, the base station must to wait approximately 19 seconds to receive the aggregated encrypted data by using S-ECEG, while it must wait just 9.5 seconds with RSAED. Furthermore, the lightweight function saves 140 bytes of memory, compared to that used in S-ECEG.

- **Communication overhead**

In RSAED, each sensor produces a ciphertext and sends the two parts to its corresponding aggregators by using the following packet format:

| TinyOS header | Src | Ciphertext part ( R or S ) | HMAC |
|:---:|:---:|:---:|:---:|
| (5) | (1) | (21) | (10) |

Figure 3: packet format

For HMAC, the output is truncated and only 10 bytes are used. Each sensor nodes uses the above data packet format, which results in 37 bytes. And hence, our scheme avoids the problems of dividing into blocks which incurs delay and using big packets which decreases the reliability of WSN. If we assume that each cluster consist of N sensor nodes including the two aggregators, the number of packets received by aggregator nodes in our scheme is N-1, N-2 packets from sensor nodes and one packet from its brother for verification purpose. In figure 6, we compare with S-ECEG where each sensor divides the ciphertext into blocks and sends each block separately.

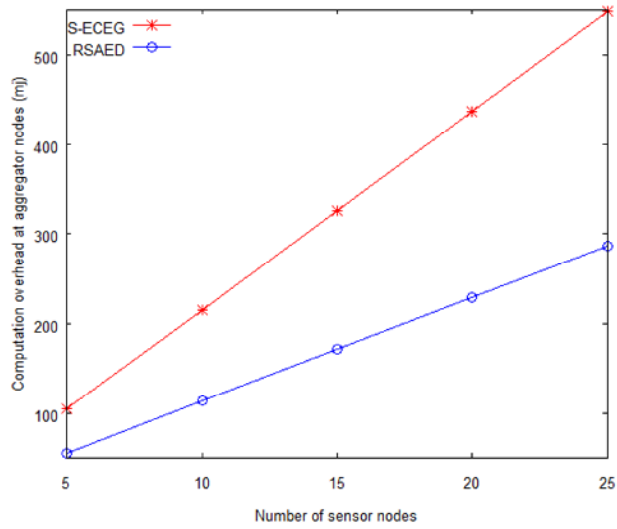

Figure 4 Computation overhead vs. Number of sensor nodes

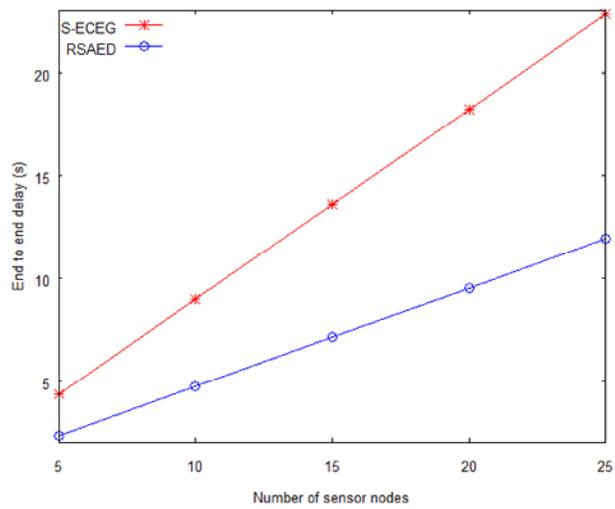

Figure 5 End-to-end delay vs. Number of sensor nodes

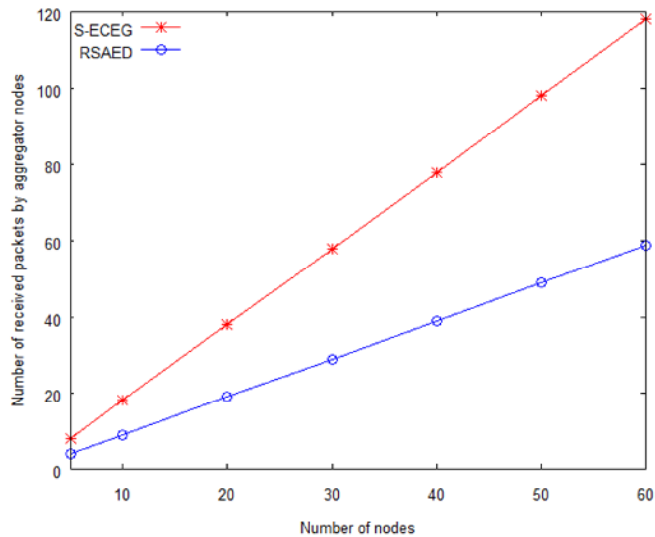

Figure 6 Number of received packets vs. Number of nodes

## 5. RELATED WORKS VS RSAED

In the area of WSN, several approaches have been proposed to secure data aggregation.

The protocol proposed in [21] provides data integrity by using delayed aggregation and delayed authentication i.e. the data aggregation is performed at grandparent nodes instead of parent nodes and the authentication is verified once the shared key is revealed by the base station. Even if the proposed scheme provides integrity, data secrecy is not provided because the data is sent in plaintext format. In our scheme, we use an additive homomorphic encryption scheme that provides the end-to-end confidentiality, where the encryption is performed at sensor node and the decryption is performed only at the base station.

The authors of [22] proposed the first secure data aggregation based on elliptic curve cryptography, in their scheme the data is encrypted hop-by-hop, and once received the CH computes the average and sends the results to its members, each sensor then compares the result with its own value and if the difference goes beyond threshold, partial signature on average is performed and sent to CH, the CH then combines all signatures in one full signature and forwards the result to the base station. Their scheme incurs high communication costs to validate the data, and can only support the average aggregation function. Our scheme reduces significantly the energy consumption due to the reduced number of messages sent to validate the data, and can support all aggregation function that can support an additive homomorphic encryption described in [5].

In [23], the authors proposed a symmetric homomorphic encryption in which an addition of plaintext to the current key shared with the base station modulo the length of key space is performed for encryption, and for decryption, the base station needs exactly the same keys used for encryption to obtain the plaintext. Their scheme provides end-to-end confidentiality, but data integrity is not addressed, furthermore, there is a problem of sensor identities because the base station must know exactly which sensor is part of aggregate to subtract the corresponding key. In our scheme we provide both data confidentiality and data integrity, and due to the use of public key encryption our scheme haven't id's problem of symmetric encryption.

The work in [24] provides both the end-to-end confidentiality and the end-to-end integrity, for the former the authors used ECEG and for integrity purpose they used aggregate signature based on bilinear maps [25], the features of an aggregate signature scheme are that the final verifier has to know not only the individual data but also the public key used for signature in order to verify the final aggregate signature, for this purpose the authors employ an encoding function that allows the base station to extract individual data, the problem in this function is that if the number of node increases the packet size increases. In our scheme the verification is performed hop-by-hop using HMAC which leads to lesser computation overhead comparing with their scheme, and also the packet size is always the same in RSAED.

In [26], the authors proposed the end-to-end integrity with aggregate signature by using a signature scheme without hash function to allow addition operation. Their scheme leads to an important computation and communication overhead because the packet includes ciphertext, signature, and public key. Also, their scheme can only verify the final aggregate, so if the verification fail, then an important number of legitimate packets is lost which leads to a significant waste of valuable network resources. In our scheme, the hop-by-hop verification saves bandwidth and computation capabilities, and allows verifier to drop forge packets, therefore, the legitimate packets can reach the base station and the robustness of WSN is improved.

## 6. CONCLUSION

We have presented RSAED to secure data aggregation in WSN. The proposed scheme is based on an additive homomorphic encryption algorithm that allows aggregation on encrypted data. In addition, our scheme provides data integrity and improves the robustness of the network. Implementation results show RSAED's applicability to WSN. In future work, we aim to improve the performance of the most expensive elliptic curve operations of our scheme namely the scalar point multiplication and the point decompression and also, provide further simulations.

## REFERENCES

[1] Jennifer Yick, Biswanath Mukherjee, Dipak Ghosal, "Wireless sensor network survey," ELSEVIER Computer Networks 52, pp. 2292–2330, 2008.

[2] K. Akkaya, M.Demirbas, R.S. Aygun, The Impact of Data Aggregationon the performance of Wireless Sensor Networks, Wiley Wireless Communication Mobile Computing (WCMC), J(8), 171-193, 2008.

[3] Xiangqian Chen, Kia Makki, Kang Yen, Niki Pissinou: Sensor network security: a survey. IEEE Communications Surveys and Tutorials 11(2): 52-73 (2009).

[4] Alzaid, Hani and Foo, Ernest and Gonzalez Nieto, Juan M. (2008) *Secure Data Aggregation in Wireless Sensor Networks: A survey*. In Proceedings Australasian Information Security Conference 2008 : Conferences in Research and Practice in Information Technology (CRPIT) 81, Wollongong, NSW, Australia.

[5] Steffen Peter, Dirk Westhoff, Claude Castelluccia: A Survey on the Encryption of Convergecast Traffic with In-Network Processing. IEEE Trans. Dependable Sec. Comput. 7(1): 20-34 (2010).

[6] Fontaine and Galand. A Survey of Homomorphic Encryption for Nonspecialists. In *EURASIP Journal on Information Security*, volume 2007, pages 1-15, 2007.

[7] J. Domingo-Ferrer. A provably secure additive and multiplicative privacy homomorphism. 6th ISC conference, pages 471–483, 2003.

[8] J. Cheon, W.-H. Kim, and H. Nam. Known-plaintext cryptanalysis of the domingo ferrer algebraic privacy homomorphism scheme. Inf. Processing Letters, 97(3):118–123, 2006.

[9] Nils Gura, Arun Patel, ArvinderpalWander, Hans Eberle, and Sheueling Chang Shantz. Comparing elliptic curve cryptography and rsa on 8-bit cpus. In CHES, Boston, pages 119–132, 2004. http://research.sun.com/projects/crypto/CHES2004.pdf.

[10] Darrel Hankerson, Alfred J. Menezes, and Scott Vanstone. Guide to Elliptic Curve Cryptography. Springer, New York [et al.], 2004.

[11] E. Mykletun, J. Girao, and D. Westhoff, "Public Key Based Cryptoschemes for Data Concealment in Wireless Sensor Networks", IEEE Int'l Conf. Communications (ICC '06), 2006. pp. 2288-2295.

[12] Certicom Research. Standards for efficient cryptography – SEC 2: Recommended elliptic curve domain parameters. http://www.secg.org/collateral/sec2_final.pdf, September 2000.

[13] Mihir Bellare, Ran Canetti et Hugo Krawczyk, Keying Hash Functions for Message Authentication, CRYPTO 1996, pp. 1–15.

[14] J. Pollard, Monte Carlo methods for index computation mod p, Mathematics of Computation, Volume 32, 1978.

[15] Crossbow Technology Inc., 2007.http://www.xbow.com.

[16] An Liu, Peng Ning, "TinyECC: Elliptic Curve Cryptography for SensorNetworks(Version1.0)",http://discovery.csc.ncsu.edu/software/TinyECC/, November 2007.

[17] http://www.tinyos.net.

[18] Avrora, The AVR simulation and analysis framework http://compilers.cs.ucla.edu/avrora/index.html.

[19] An Liu, Peng Ning, "TinyECC: A Configurable Library for Elliptic Curve Cryptography in Wireless Sensor Networks," in Proceedings of the 7th International Conference on Information Processing in Sensor Networks (IPSN 2008), SPOTS Track, pages 245--256, April 2008.

[20] Roberto M. Avanzi, Henri Cohen, Christophe Doche, Gerhard Frey, Tanja Lange, Kim Nguyen, Frederik Vercauteren "Handbook of Elliptic and Hyperelliptic Curve Cryptography" (2006).

[21] L. Hu, D. Evans, Secure aggregation for wireless networks, Proceedingsof the Workshop on Security and Assurance in Ad Hoc Networks, Orlando,FL, 28 January 2003.

[22] Mahimkar, A., Rappaport, T.S.: SecureDAV: A secure data aggregation and verification protocol for sensor networks. In: Proceedings of the IEEE Global Telecommunications Conference (2004).

[23] Castelluccia, C., Mykletun, E., Tsudik, G.: Efficient Aggregation of Encrypted Data in Wireless Sensor Networks. In: MobiQuitous (2005).

[24] Sun, H.-M., Hsiao, Y.-C., Lin, Y.-H., Chen, C.-M.: An Efficient and Verifiable concealed Data Aggregation Scheme in Wireless Sensor Networks. In: Proceedings of the 2008 International Conference on Embedded Software and Systems, pp. 19–26 (2008).

[25] D. Boneh, C. Gentry, B. Lynn, and H. Shacham, "Aggregate and verifiably encrypted signatures from bilinear maps", In Proc. EUROCRYPT 2003, pp. 416-432. 2003.

[26] Albath, J., Madria, S.: Secure Hierarchical Aggregation in Sensor Networks. In: Proceedings of IEEE Wireless Communications and Networking Conference (2009).

**AUTHORS**

**Merad Boudia Omar Rafik** received his Master degree in Computer Science from the University of Tlemcen (Algeria) in 2010. Member of STIC laboratory in the same university, his research interests include sensor network security and cryptography.

**Mohammed Feham** received his PhD in Engineering in optical and microwave communications from the University of Limoges (France) in 1987, and his PhD in science from the university of Tlemcen (Algeria) in1996. Since 1987, he has been Assistant Professor and Professor of Microwave, Communication Engineering and Telecommunication network. His research interests cover telecommunication systems and mobile networks.